\documentclass[%
preprint,
nofootinbib,
 amsmath,amssymb,
 aps,
]{revtex4-2}

\usepackage{graphicx}% Include figure files
\usepackage{dcolumn}% Align table columns on decimal point
\usepackage{bm}% bold math
\begin{document}

\def\bb    #1{\hbox{\boldmath${#1}$}}

\title{ Probing multi-particle bunching from intermittency analysis \\ in relativistic heavy-ion collisions}

\author{Valeria Zelina Reyna Ortiz} 
\email{val.reyna@cern.ch}
\affiliation{Institute of Physics, Jan Kochanowski University, 25-406 Kielce, Poland}
\author{Maciej Rybczy\'nski}
\email{maciej.rybczynski@ujk.edu.pl}
\affiliation{Institute of Physics, Jan Kochanowski University, 25-406 Kielce, Poland}
\author{Zbigniew W\l odarczyk}
\email{zbigniew.wlodarczyk@ujk.edu.pl}
\affiliation{Institute of Physics, Jan Kochanowski University, 25-406 Kielce, Poland}

\begin{abstract}
It has been demonstrated that factorial moments analysis in dependence from the size of phase-space cells (when the latter is decreased but is still considerably large), exhibits sensitivity to particle bunching within a system situated in the two-dimensional cell. Based on recent findings concerning fluctuations in charge particle density observed in central Au + Au collisions at various energies ranging from $\sqrt{s_{NN}}$ = 7.7 to 200 GeV at RHIC/STAR, it is proposed that analyzing factorial moments in relativistic heavy-ion collisions could serve as a method to investigate density fluctuations linked to correlation phenomena. Our understanding of particle production mechanisms may be enriched by using factorial moments analysis to gain new information contained in them that was previously unavailable.
\end{abstract}

\pacs{XXX}

\maketitle
\section{Introduction}
\label{Introduction}

The major goal of the intermittency (a self-similar correlation regarding the size of the phase-space volume) analysis in relativistic heavy ion collision programs is to explore the phase diagram of quantum chromodynamic (QCD) matter. The critical end point (CEP) is a key feature of the QCD phase diagram, representing the point where the first-order phase transition boundary terminates~\cite{Bowman:2008kc, Hatta:2003wn}. Many efforts have been made to search for the possible CEP in heavy-ion collisions~\cite{Bzdak:2019pkr, Luo}. 

Recently, there has been discussion regarding the potential of particle density fluctuations in heavy-ion collisions to serve as a distinctive indicator of the phase transition within the QCD phase diagram~\cite{Li:2015pbv, Sun:2018jhg}. Critical opalescence, a notable light scattering phenomenon observed in continuous (or second-order) phase transitions in conventional quantum electrodynamic matter~\cite{Brumberger, Lesne}, is characterized by significant fluctuations within the critical region. In the context of heavy-ion collisions, akin to conventional critical opalescence, it is anticipated that the produced matter would exhibit pronounced particle density fluctuations close to the CEP due to the rapid increase in correlation length within the critical region~\cite{Stephanov:1998dy, Berdnikov:1999ph}. Such large density fluctuations are expected to persist through the kinetic freeze-out stage if they can withstand final-state interactions during the hadronic evolution of the system.

Upon approaching a critical point, the correlation length of the system diverges and the system becomes scale invariant, or self-similar~\cite{DeWolf:1995nyp, Bialas:1990xd, Satz:1989vj}. Based on the 3D-Ising universality class arguments~\cite{Antoniou:2006zb, Antoniou:2017vti, Antoniou:2000ms, Antoniou:2015lwa}, the
density-density correlation function for small momentum transfer has a power-law structure, leading to large density fluctuations in heavy-ion collisions~\cite{Antoniou:2006zb, Antoniou:2017vti, Antoniou:2000ms, Antoniou:2015lwa}. Such fluctuations can be probed in transverse momentum phase space within the framework of
an intermittency analysis by utilizing the scaled factorial moments~\cite{Antoniou:2006zb, Antoniou:2017vti, Antoniou:2000ms, Antoniou:2015lwa, NA49:2009diu, Wu:2019mqq}. The methodology  consists of dividing the D-dimensional phase space into $M^D$ equal-sized cells and the observable, $q$-th order scaled factorial moment  
$F_{q}\left(M\right)$, is defined as follows~\cite{Antoniou:2006zb, Hwa:1992uq, NA49:2012ebu, Antoniou:2005am, Bialas:1985jb, Bialas:1988wc}:
\begin{equation}
F_{q}\left(M\right) = \frac{\langle \frac{1}{M^{D}}\sum_{i=1}^{M^{D}} n_{i}\left(n_{i}-1\right)\ldots\left(n_{i}-q+1\right) \rangle}{\langle \frac{1}{M^{D}} \sum_{i=1}^{M^{D}} n_{i}  \rangle^{q}}, 
\label{Fq}
\end{equation}
where $M^{D}$ is the number of cells in D-dimensional phase space and $n_{i}$ is the measured multiplicity of a given event in the $i$-th cell. The angle bracket denotes an average over the events. The intermittency appears as a power-law (scaling) behavior of scaled factorial moments, and factorial moments increase when the size of the phase-space cells goes to zero\footnote{Frequently in multiparticle production, a system is said to be "intermittent" if a power-law relations of factorial moments exist~\cite{STAR:2023jpm}. Usually one observes an increase of factorial moments with decreasing size of phase-space cells but this happens for quite large cells and not in the limit where the size of phase-space cells goes to zero. However, for smaller cells the factorial moments saturate and no intermittency effect can be observed.}~\cite{Antoniou:2006zb, Hwa:1992uq, Antoniou:2000ms, Bialas:1985jb}. If the system features density fluctuations, scaled factorial moments will obey a power-law behavior of $ F_{q}\left(M\right) \propto \left(M^{D}\right)^{\phi_{q}}$, $ M\gg 1$, where $\phi_{q}$ is called the intermittency index quantifying the strength of intermittency~\cite{Antoniou:2006zb, Antoniou:2017vti, Antoniou:2000ms, NA49:2009diu, NA49:2012ebu}. 

However, the consideration of "intermittency" as a genuinely novel phenomenon remained less apparent. Relation between Hanbury-Brown-Twiss effect and intermitency-like phenomenon was discussed and advocate that they are perfectly compatible~\cite{Bialas:1992ca, Wibig:1995jg}. Has been shown that the multi-particle Bose-Einstein correlations leads to the power-law like behavior of scaled factorial moments. In Ref.~\cite{Charlet:1993ic}, diverse data sets were analyzed, leading to the conclusion that intermittency results from Bose-Einstein correlations, alongside a mechanism driving power-law behavior. Conversely, in Ref.~\cite{Carruthers:1989jj}, the authors contend that the observed intermittent-like patterns in moments of multiplicity distributions are attributable to the quantum statistical properties of the particle-emitting system, rather than serving as direct evidence of intermittency.

Notice that implementing the "mixed event method" effectively removes contributions of multiplicity fluctuations (among others background contributions)~\cite{NA49:2009diu, NA49:2012ebu, Antoniou:2005am}. Mixed events are constructed by randomly selecting particles from different original events, while ensuring that the mixed events have the same multiplicity and momentum distributions as the original events. Usually instead of $F_q(M)$ the following observable is used:
\begin{equation}
\Delta F_{q}\left(M\right) =F_{q}\left(M\right)^{data}-F_{q}\left(M\right)^{mix}, 
 \label{deltaFq}
\end{equation}
where the moments from mixed events are subtracted from the data.

Recently the STAR Collaboration published extensive data on $\Delta F_{q}\left(M\right)$ in two dimensional ($D=2$) transverse momentum plane ($p_{Tx},p_{Ty}$) from central Au+Au collisions at $\sqrt{s_{NN}}$ = 7.7, 11.5, 19.6, 27, 39, 62.4 and 200 GeV~\cite{STAR:2023jpm}. In the next sections, based on the STAR experiment data, we provide a discussion concerning multi-particle bunching and suggest that measuring the scaled factorial moments in relativistic heavy-ion collisions could serve as a means to investigate density fluctuations within a highly constrained number of cells.  This observation will be our main point for further discussion and calculations described in next sections. 

The 0-5~$\%$ most central Au+Au collisions at $\sqrt{s_{NN}}=7.7$~GeV are discussed in Sections~\ref{one bunch} and~\ref{many bunches}, while their energy dependence is considered in Section~\ref{energy}. In Section~\ref{sum-res} the scale-invariance as a possible origin of power-law behavior of $\Delta F_{q}\left(M\right)$ is discussed. Section~\ref{Concl} summarizes and concludes our work.

\section{Multiplicity distributions of particles in cells}
\label{one bunch}

For large number of cells $M^2\gg N$ (where $ N=\sum n_{i}$), the average multiplicity in a cell $\langle n_{i}\rangle$ is small and usually one observes zero or one particle in a given cell. In such a case the multiplicity distributions of particles in a cell are indistinguishable. Considering two borderline multiplicity distributions\footnote{For $n$-particle correlation function $g_{\left(n\right)}=\left(n-1\right)!$ we have Bose-Einstein distribution, while for Poisson distribution $g_{\left(n\right)}=0$. Between these distributions we have negative binomial distribution and $g_{\left(n\right)}=\left(n-1\right)!k^{-n+1}$ with real positive parameter $k$. The $n!$ dependence is a consequence of Wick's theorem~\cite{Wick} and its validity was first demonstrated with thermal photons~\cite{Glauber} and recently proved for massive particles~\cite{Dall}.}, Poisson distribution:
\begin{equation}
P\left(n_i\right)=\frac{\langle n_{i}\rangle^{n_{i}}}{n_{i}!}\exp\left(-\langle n_{i}\rangle\right) 
\label{PD}
\end{equation}
and geometric (Bose-Einstein) distribution:
\begin{equation}
P\left(n_{i}\right)=\frac{1}{1+\langle n_{i}\rangle}\left(\frac{\langle n_{i}\rangle}{1+\langle n_{i}\rangle}\right)^{n_{i}},
\label{BE}
\end{equation}
one can see that for small $\langle n_{i}\rangle$, $P\left(0\right)=\exp\left(-\langle n_{i}\rangle\right)\simeq 1-\langle n_{i}\rangle$ and $P\left(1\right)=\langle n_{i}\rangle\exp\left(\langle n_{i}\rangle\right)\simeq \langle n_{i}\rangle$ for Poisson distribution, and similarly $P\left(0\right)=1/\left(1+\langle n_{i}\rangle\right) \simeq 1-\langle n_{i}\rangle$ and $P\left(1\right)=\langle n_{i}\rangle/\left(1+\langle n_{i}\rangle\right)^{2} \simeq \langle n_{i}\rangle$ for geometric distribution~\cite{Goodman}. 
\begin{figure}
\begin{center}
\includegraphics[scale=0.48]{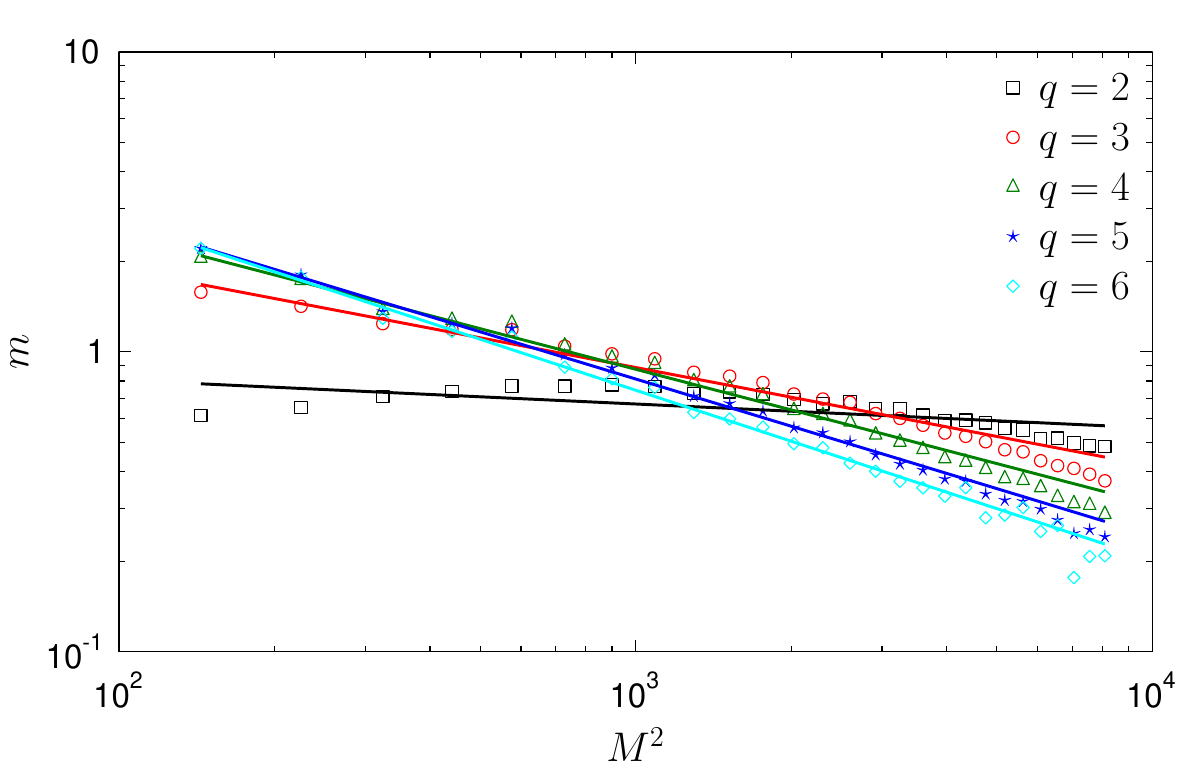}
\end{center}
\vspace{-5mm}
\caption{Average multiplicity of charged particles in a bunch of particles as a function of the number of cells $M^2$ evaluated from $\Delta F_{q}$ for $\sqrt{s_{NN}}=7.7$~GeV.}
\label{SPD-1}
\end{figure}
\begin{figure}
\begin{center}
\includegraphics[scale=0.48]{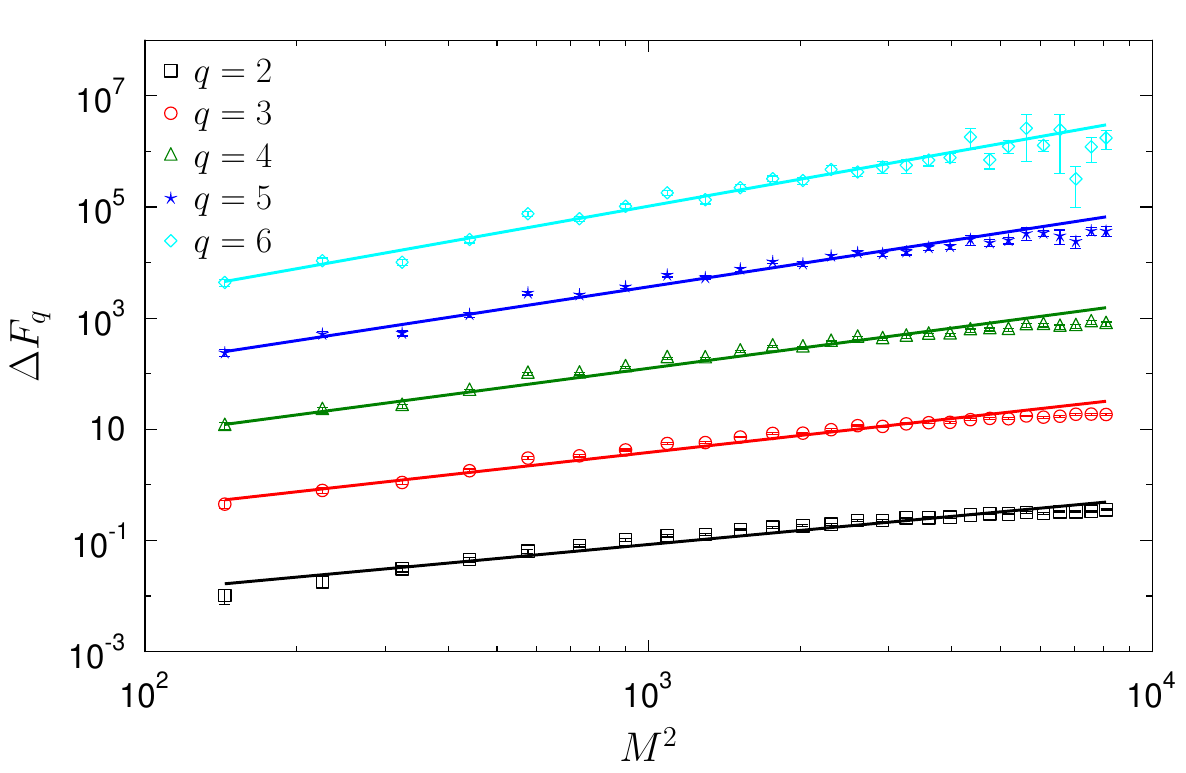}
\end{center}
\vspace{-5mm}
\caption{$\Delta F_{q}$ evaluated for corresponding average multiplicity of particles in a bunch, in comparison with experimental data~\cite{STAR:2023jpm}.}
\label{SPD-4}
\end{figure}

\begin{figure}
\begin{center}
\includegraphics[scale=0.48]{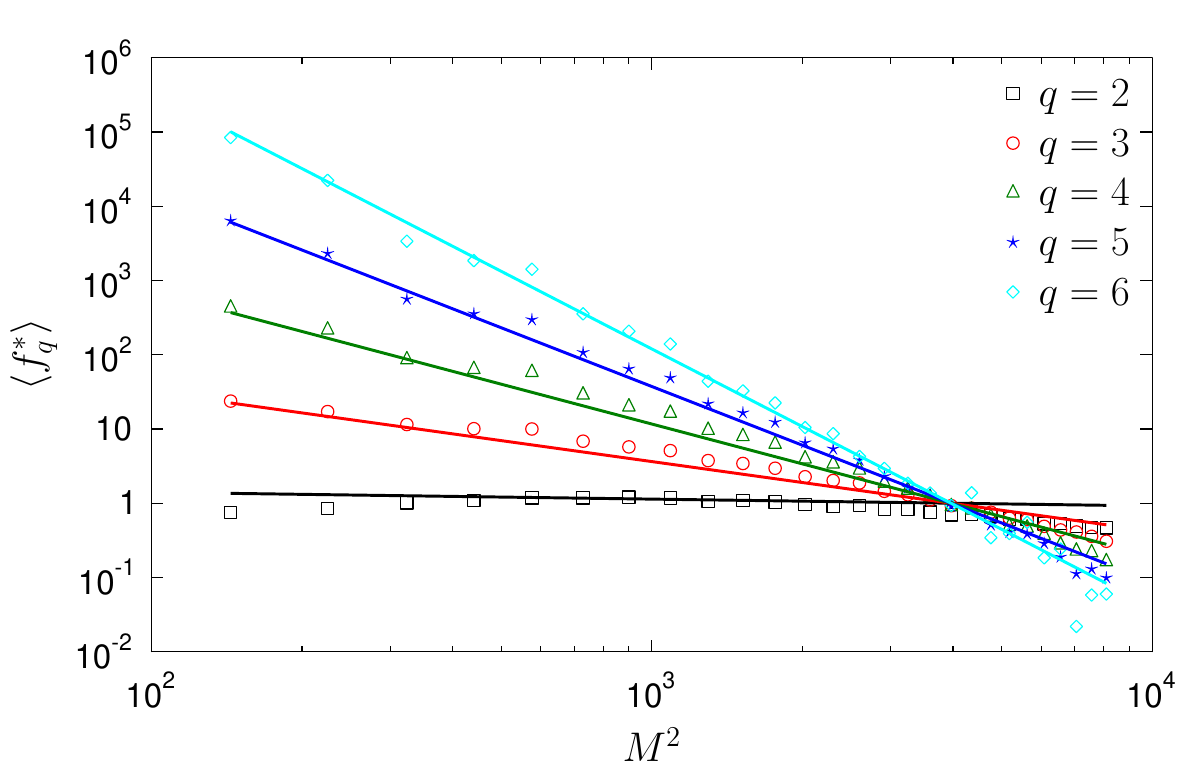}
\end{center}
\vspace{-5mm}
\caption{Average factorial moments $\langle f_{q}^{*}\rangle$ evaluated from experimental $\Delta F_{q}$ at $\sqrt{s_{NN}}=7.7$~GeV. 
Fits show dependence of $\langle f_{q}^{*}\rangle = \left(M^{2}/3962\right)^{b} $ with $b=1.6-0.846\cdot q$.}
\label{SPD-2}
\end{figure}

\begin{figure}
\begin{center}
\includegraphics[scale=0.48]{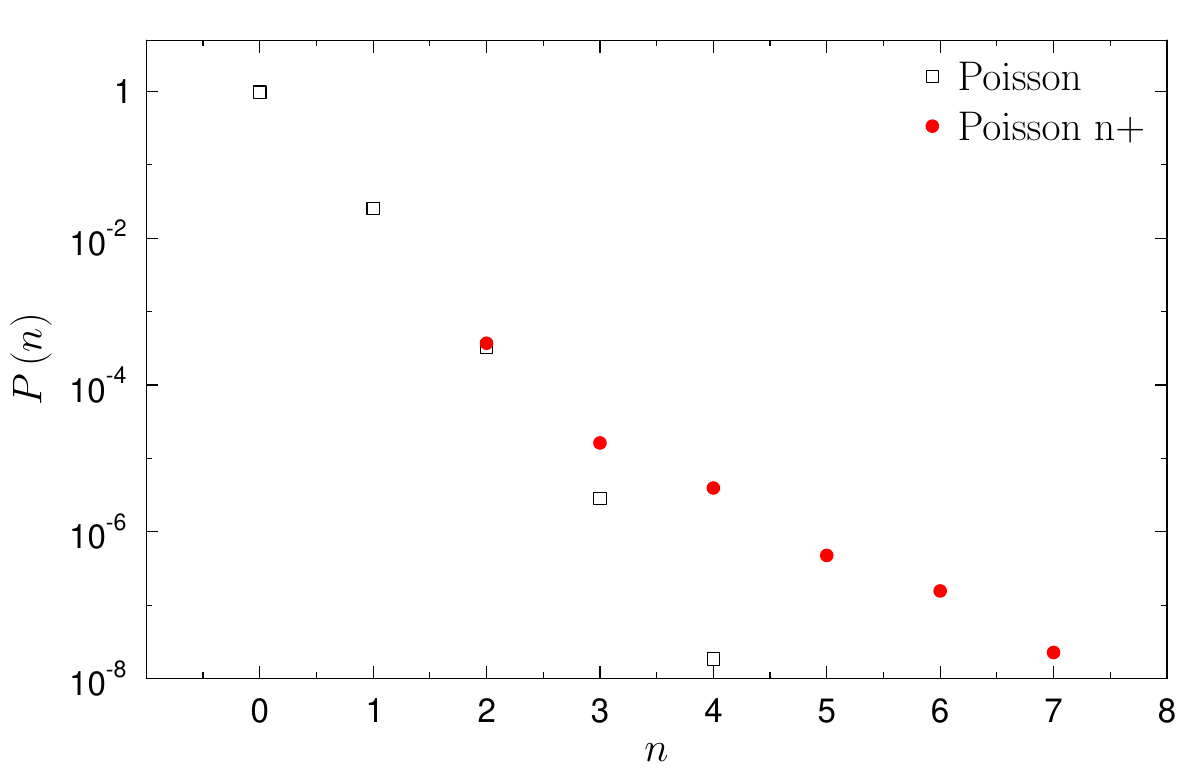}
\end{center}
\vspace{-5mm}
\caption{Multiplicity distribution of particles in one cell ($M^{2}=3962$ and $\sqrt{s_{NN}}=7.7$~GeV).}
\label{SPD-3}
\end{figure}

Assuming that Bose-Einstein correlated bunch of particles appears only in one cell, one have (for large number of cells): $F_{q} \simeq F_{q}^{PD}+\langle f_q^{*}\rangle\left(\langle n_{i}\rangle^{q} M^{2}\right)^{-1}$ and finally
\begin{equation}
\Delta F_{q} \simeq \langle f_{q}^{*}\rangle \frac{\left(M^{2}\right)^{q-1}}{\langle N\rangle^{q}},  
\label{DFq-1}
\end{equation}
where $\langle N\rangle=\sum n_{i}$ is the average multiplicity for $M^{2}=1$ and
\begin{equation}
\langle f_{q}^{*}\rangle = \langle n\left(n-1\right)\ldots\left(n-q+1\right)\rangle = \left<\frac{n!}{\left(n-q\right)!}\right> 
\label{fq-star}
\end{equation}
is the factorial moment averaged over multiplicity distribution in a cell with bunch of particles. For geometric distribution (\ref{BE}) with the average multiplicity $m$ one have $f_{q}^{*}=m^{q} q!$, and from Eq.~(\ref{DFq-1}) one can evaluate
\begin{equation}
m = \langle N\rangle\left(\frac{\Delta F_{q}}{\left(M^2\right)^{q-1}q!}\right)^{1/q}.
\label{mean}
\end{equation}
Fig.~\ref{SPD-1} shows average multiplicity in a bunch evaluated from experimental $\Delta F_{q}$ values. Roughly $m=a\left(M^{2}/1000\right)^{b}$ with $a \simeq 3.32\cdot  q^{-0.732}-5.33\cdot q^{-2}$ and $b=1.46\cdot q^{-1}-0.81$. Fig.~\ref{SPD-4} shows corresponding $\Delta F_{q}$ values compared to experimental data. Without assumption on multiplicity distribution of particle's bunch one can evaluate factorial moments $\langle f_{q}^{*}\rangle$ (see Fig.~\ref{SPD-2}) 
% and in which $<f_q^*>$ for $M^2$ =3962 at $\sqrt{s_{NN}} =7.7$ GeV are shown) 
and estimate multiplicity distribution $P\left(n\right)$ in a bunch. From Eq.~(\ref{fq-star}) one can write
\begin{equation}
\langle f_{q}^{*}\rangle = \sum_{n=q}^{n_{max}} \frac{n!}{\left(n-q\right)!} P\left(n\right)    
\label{fq-star-2}
\end{equation}
and
\begin{equation}
P\left(n=q\right) = \frac{1}{q!}\left(\langle f_{q}^{*}\rangle-\sum_{i=q+1}^{n_{max}}\frac{i!}{(i-q)!} P\left(i\right)\right),
\label{Pn}
\end{equation}
where we put in calculations $n_{max}=q_{max}+1$ and $P\left(n_{max}\right)\simeq 0.1\cdot P\left(n_{max}-1\right)$. Estimated multiplicity distributions in cells (for $\langle f_{q}^{*}\rangle$ taken at $M^{2}=3962$ and $\sqrt{s_{NN}}=7.7$~GeV) is presented in Fig.~\ref{SPD-3}. In one cell we have particle bunch with $P\left(n\right)$ given by Eq.~(\ref{Pn}) and multiplicity distributions in all the other cells are given by Poisson distribution (\ref{PD}) with $\langle n_{i}\rangle=N/M^{2}$.

\section{Many bunches case}
\label{many bunches}


\begin{figure}
\begin{center}
\includegraphics[scale=0.48]{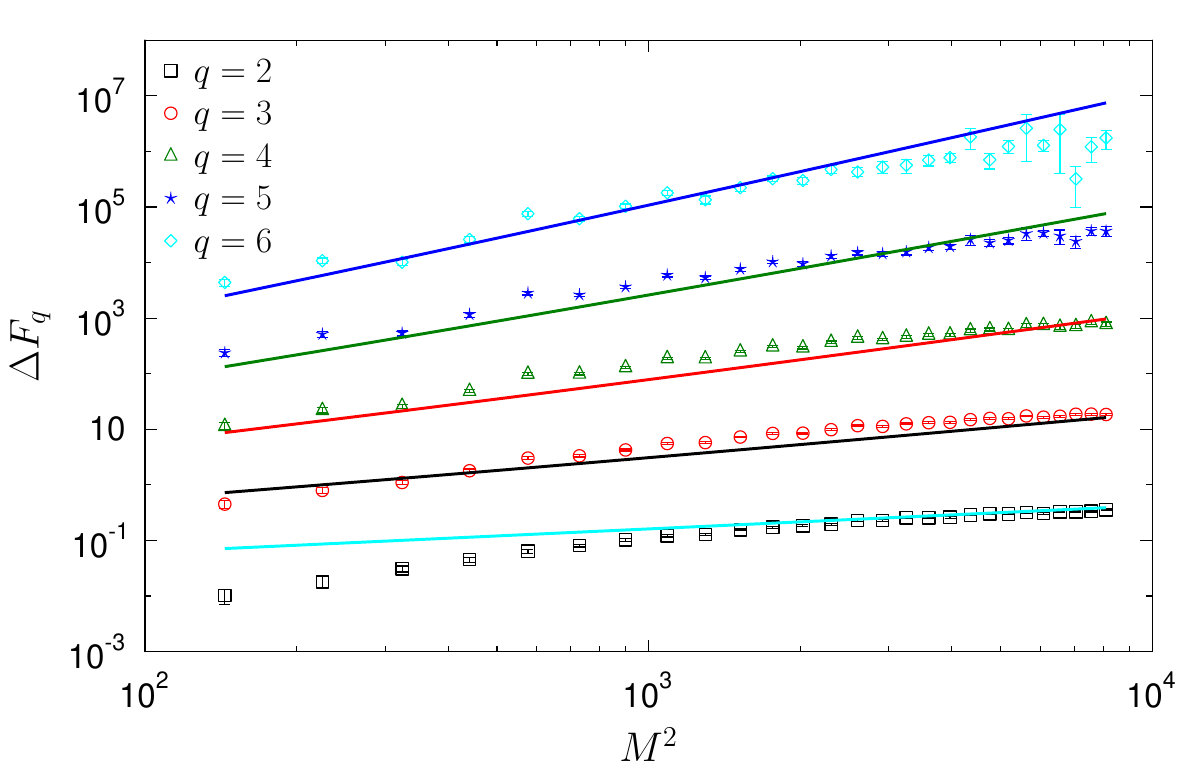}
\end{center}
\vspace{-5mm}
\caption{ $\Delta F_{q}$ from Eq.~(\ref{DFq-2}) in comparison with experimental data at $\sqrt{s_{NN}}=7.7$~GeV.}
\label{SPD-5}
\end{figure}

For the case with $k$ bunches with multiplicity governed by geometric distribution~(\ref{BE}) one have negative binomial distribution (NBD)
\begin{equation}
P\left(n\right)= \frac{\Gamma\left(n+k\right)}{\Gamma\left(n+1\right) \Gamma\left(k\right)} \left(1+\frac{k}{\langle n\rangle}\right)^{-n}\left(1+\frac{\langle n\rangle}{k}\right)^{-k}
\label{NBD}
\end{equation}
and factorial moments
\begin{equation}
\langle f_{q}\rangle = \frac{\Gamma\left(k+q\right)}{\Gamma\left(k\right)}\frac{\langle n\rangle^{q}}{k^{q}}.
 \label{fq-NBD}
\end{equation}
Correlations being a signature of Bose-Einstein statistics never reaches its maximum allowed degree, the most natural interpretation is that the source is not totally chaotic as would be expected from a purely thermalized system, but that a certain degree of coherence ($1-\lambda$) is present in the boson radiation~\cite{Fowler:1977gx}. Effective factorial moments are expected to be
\begin{equation}
\langle f_{q}\rangle=\left(1-\lambda\right)\langle f_{q}^{PD}\rangle+\lambda\langle f_{q}^{NBD}\rangle,  
 \label{fq-total}
\end{equation}
where $\langle f^{NBD}\rangle$ (c.f. Eq.~(\ref{fq-NBD})) and $\langle f^{PD}\rangle=\langle n\rangle^{q}$ are factorial moments for negative binomial and Poisson distributions, respectively.
In this case, one have 
\begin{equation}
\Delta F_{q}\simeq\lambda\frac{\Gamma\left(k+q\right)}{\Gamma\left(k\right)k^{q}}-\lambda.
 \label{DFq-2}
\end{equation}
Expected $\Delta F_{q}$ in comparison with experimental data are shown in Fig.~\ref{SPD-5}. Comparing Eq.~(\ref{DFq-2}) with experimental values at $\sqrt{s_{NN}}=7.7$~GeV, we estimate roughly $\lambda=0.02$ and $k=2.56 \left(M^{2}\right)^{-0.42}$. 
For small $k$-parameter, Eq.~(\ref{DFq-2}) leads to relation $\Delta F_{q} \propto k^{1-q}$ and for evaluated $k$-parameter we have dependence
\begin{equation}
\Delta F_{q} \propto \left(M^{2}\right)^{0.42\left(q-1\right)}.
 \label{DFq-3}
\end{equation}

\section{Energy dependence}
\label{energy}

\begin{figure}
\begin{center}
\includegraphics[scale=0.48]{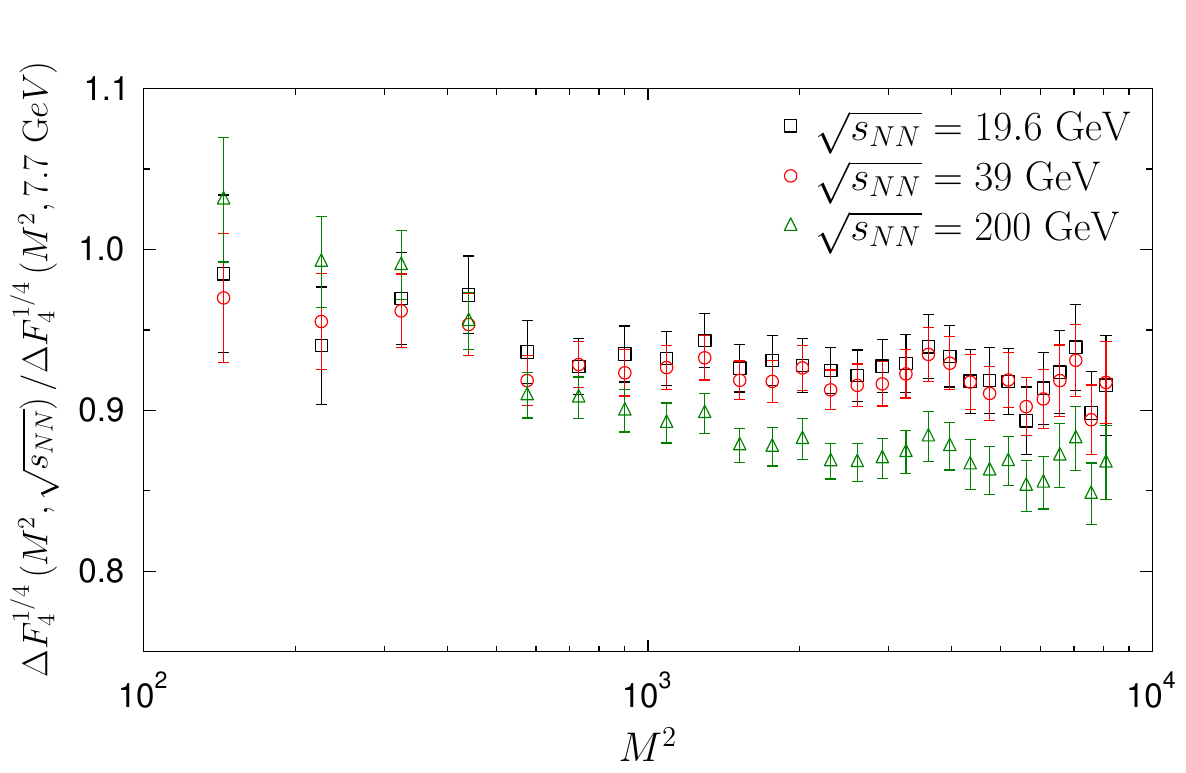}
\end{center}
\vspace{-5mm}
\caption{Ratios $\Delta F_{4}^{1/4}\left(M^{2}, \sqrt{s_{NN}}\right)/\Delta F_{4}^{1/4}\left(M^{2}, 7.7~{\textrm GeV}\right)$ for different $\sqrt{s_{NN}}$ energies.} 
\label{SPD-6}
\end{figure}

\begin{figure}
\begin{center}
\includegraphics[scale=0.48]{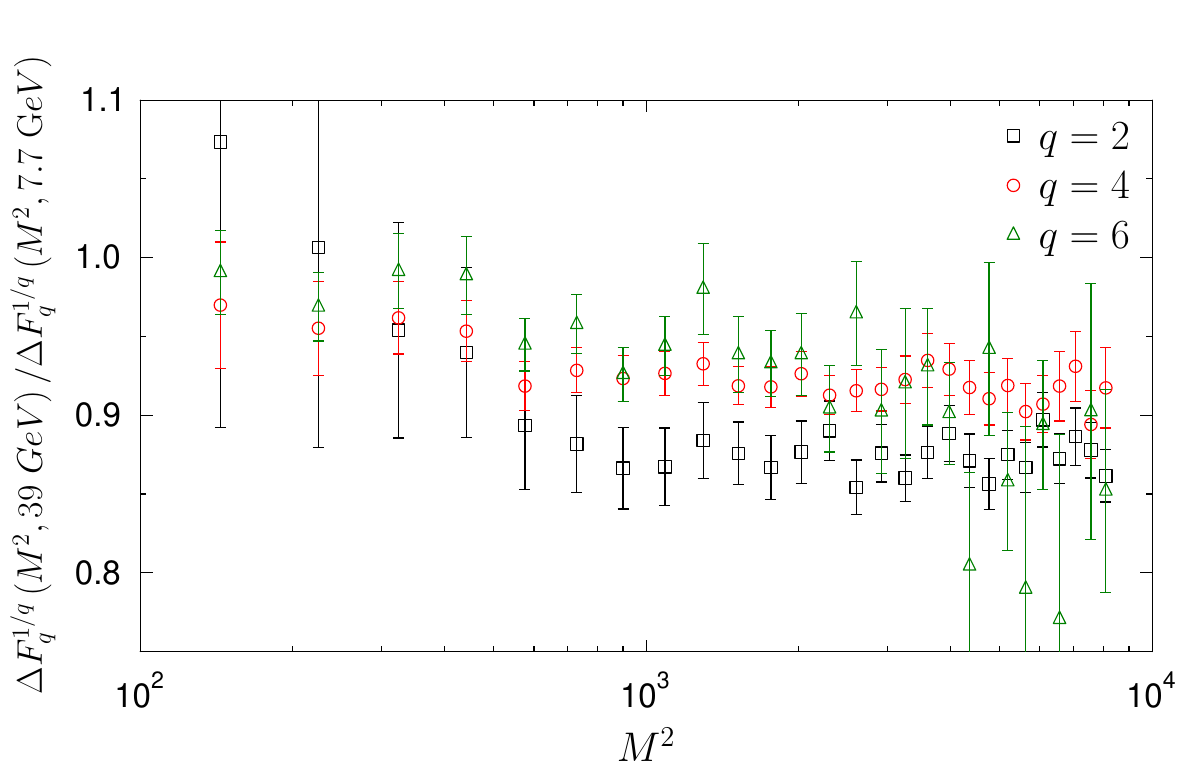}
\end{center}
\vspace{-5mm}
\caption{Ratios $\Delta F_{q}^{1/q}\left(M^2, 39~{\textrm GeV}\right)/\Delta F_q^{1/q}\left(M^2,7.7~{\textrm GeV}\right)$ for different $q$ values.}
\label{SPD-7}
\end{figure}

Scaled factorial moments are weekly dependent on collision energy. In Figs.~\ref{SPD-6} and \ref{SPD-7} we show ratios $\Delta F_{4}^{1/4}\left(M^{2},\sqrt{s_{NN}}\right)/\Delta F_{4}^{1/4}\left(M^{2},7.7~{\textrm GeV}\right)$ \newline and $\Delta F_{q}^{1/q}\left(M^{2}, \sqrt{s_{NN}}=39~{\textrm GeV}\right)/\Delta F_q^{1/q}\left(M^{2},7.7~{\textrm GeV}\right)$, respectively, for different energies $\sqrt{s_{NN}}$ and different values of order $q$. Taking into account Eq.~(\ref{mean}), this ratio corresponds with ratios $m/\langle N\rangle$ at different energies. For large $M^{2} > 1000$, $\frac{m(\sqrt{s_{NN}})}{\langle N(\sqrt{s_{NN}}\rangle)}=\frac{m(7.7~GeV)}{\langle N(7.7~{\textrm GeV})\rangle}(0.985-0.0236 \ln{\sqrt{s_{NN}}})$. Weak energy dependence indicate that energy dependence $m(\sqrt{s_{NN}})$ follows energy dependence of total multiplicity $\langle N(\sqrt{s_{NN}})\rangle$. 

\section{Possible origin of power-law behavior}
\label{sum-res}

For a scenario in which a bunch of particles appears only in one cell (or in a limiting number of cells), following Eq.~(\ref {DFq-1}), we expect a power-law behavior of scaled factorial moments, $\Delta F_{q} \propto (M^{2})^{\Phi_{q}}$ with the power index $\Phi_{q}=q-1$, for multiplicity $m$ of particles in a bunch being independent on the number of cells $M^{2}$. In reality, the division procedure of phase space changes multiplicity $m$ in a cell. Number of particles in bunch of effective space size $\delta$ in cell of size $\Delta\left(M\right)=\Delta\left(M=1\right)/M$ is expected to be $m=m(\delta/\Delta(M))=m(M)$.

\begin{figure}
\begin {center}
\includegraphics[scale=0.48]{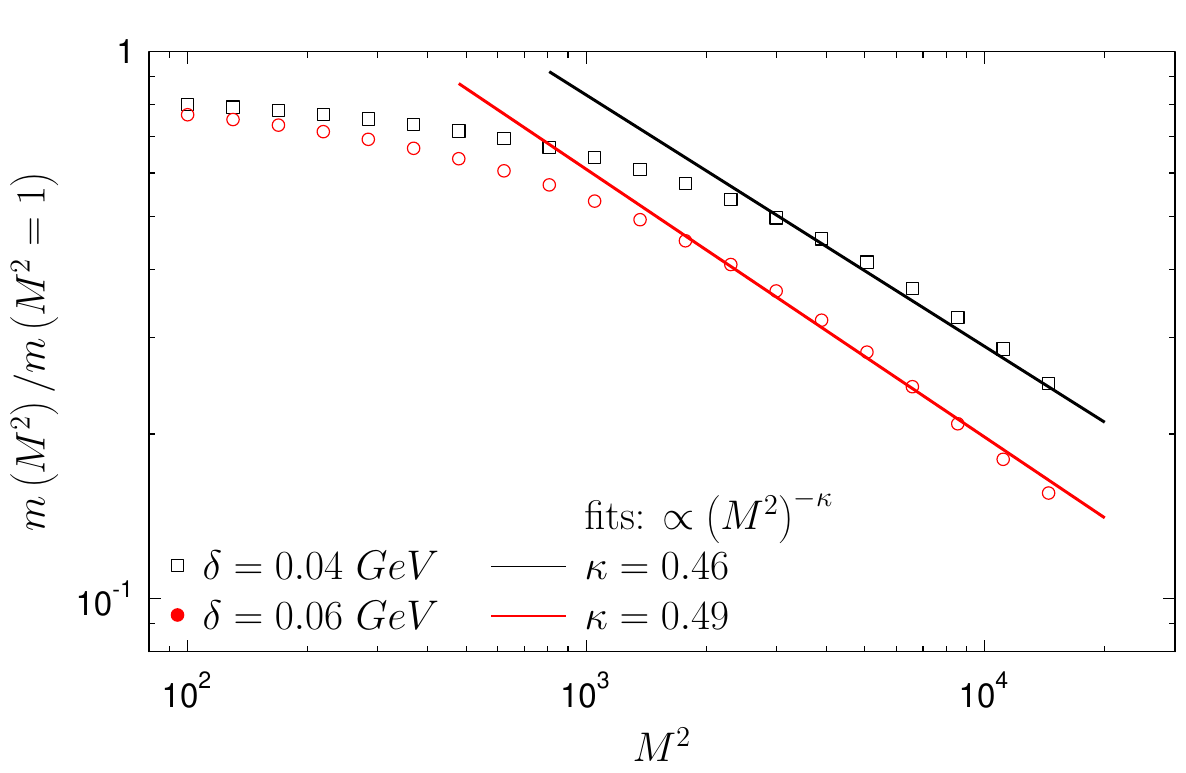}
\end{center}
\vspace{-5mm}
\caption{ The relative multiplicity in bunch $m\left(M^{2}\right)/m\left(M^{2}=1\right)$ dependence on number of cells $M^{2}$ for space size $\delta=0.04$~GeV and $\delta=0.06$~GeV. For large $M^{2}$ we observe a power-law dependence $m\left(M^{2}\right) \propto \left(M^{2}\right)^{- \kappa}$ with $\kappa$=0.46 and 0.49, respectively.} 
\label{SPD-8}
\end{figure}

Let's consider a bunch of particles of transverse size $\delta$ (dispersion with respect to the axis of a bundle) in the random direction $(p_{Tx0},p_{Ty0})$ with a 
transverse momentum distribution
\begin{equation}
f(p_{Tx},p_{Ty})=\frac{3}{\pi \delta^{2}} \exp\left(- \frac{\sqrt{6}\sqrt{(p_{Tx}-p_{Tx0})^{2}+(p_{Ty}-p_{Ty0})^{2}}}{\delta} \right).
\label{pt}
\end{equation}
Numerical results of the $m\left(M^{2}\right)$ dependence are shown in Fig.~\ref{SPD-8}. For large $M^{2}$ we observe power-law dependence\footnote{The procedure of division of particles in a given bunch occurs as a statistical process at a deformed exponential rate. This means that the number of particles in a bunch that are likely to be removed from a cell within a given cell number range is proportional to the number of particles in a cell, $dm/d\left(M^{2}\right)=-m/\sigma^{2}$. In the considered cell division procedure, the division rate $\sigma^{2}=\sigma^{2}\left(M^{2}\right)$ depends on the numbers of cells, $M^{2}$ (for a large number of cells, the size of a single cell is small and there is more probable that the particles in a bunch are separated between different cells). For $\sigma^{2}\left(M^{2}\right)=\sigma_{0}^{2}+M^{2}/\kappa$ one has solution in the form of kappa-deformed exponential function (exponential function appears in the limit $\kappa\rightarrow\infty$), $m\left(M^{2}\right)\propto [1+M^{2}/\left(\sigma_{0}^{2}\kappa\right)]^{-\kappa}$. For large number of cells, $M^{2}\gg \sigma_{0}^{2}\kappa$, one has scale-free power-law dependence given by Eq.~(\ref{mdep}).}:
\begin{equation}
m\left(M^{2}\right) \propto \left(M^{2}\right)^{- \kappa} 
\label{mdep}
\end{equation}
which will result in
\begin{equation}
\Delta F_{q}\left(M^{2}\right) \propto m^{q}\left(M^{2}\right)\left(M^{2}\right)^{q-1}=\left(M^{2}\right)^{q\left(1-\kappa\right)-1}. 
\label{DFq-kappa}
\end{equation}

\vspace{0.6cm}

The power index $\Phi_{q}=q\left(1-\kappa\right)-1$ is the result of the division procedure, which leads to power-law dependence $m\left(M^{2}\right)$ given by Eq.~(\ref{mdep}). It is known~\cite{Sornette} that, if function $m\left(M^{2}\right)$ follows a simple power law, $m\left(M^{2}\right) \propto \left(M^{2}\right)^{-\kappa}$ then it is scale invariant and
\begin{equation}
m\left(\lambda M^{2}\right)=\mu m\left(M^{2}\right) 
\label{scale-inv}
\end{equation}
with relation $\kappa = -\ln \mu /\ln \lambda$. Selecting different functions $f(p_{Tx},p_{Ty})$ changes details but the scale invariant behavior is still visible.

\section{Conclusions}
\label{Concl}

Density fluctuations in cells lead to the power-law-like behavior of scaled factorial moments. Correlated bunch of particles which appears only in one cell can be responsible for a "intermittent-like" behavior of $\Delta F_{q}\left(M^{2}\right)$. Multiplicity distribution $P\left(m\right)$ in a bunch of particles and spatial distribution $f(p_{Tx},p_{Ty})$ determine behavior of scaled factorial moments. For small size of cell $\Delta$, multiplicity in a bunch with size $\delta$ is scale invariant (its form follows a simple power law $m\left(\Delta^{2}\right) \propto \left(\Delta^{2}\right)^{\kappa}$) which will result in $\Delta F_{q}\left(M^{2}\right) \propto \left(M^{2}\right)^{q\left(1-\kappa\right)-1}$. Whether such a simple scenario is related to the broad spectrum of experimental data further calculations from dynamical modeling of heavy-ion collisions are required.

\vspace*{0.3cm}
\centerline{\bf Acknowledgments}
\vspace*{0.3cm}
This research was supported by the Polish National Science Centre (NCN) Grant 2020/39/O/ST2/00277. In preparation of this work we used the resources of the Center for Computation and Computational Modeling of the Faculty of Exact and Natural Sciences of the Jan Kochanowski University of Kielce.

%\appendix

%\section{XXX}
%\label{FqKq}

%\section{XXX}
%\label{2BD}

\end{document}